%% file: trigger_box_article.tex
\documentclass[review,number,sort&compress]{elsarticle}

\usepackage{graphics}
\usepackage{color}
\usepackage{epsfig}
\usepackage{eso-pic}
\usepackage{subfig} 
\usepackage{xspace}

\usepackage{amsmath,amssymb,wasysym}

\usepackage[bookmarksopen=true,hyperindex=true,breaklinks=true,plainpages=false]{hyperref}

\input{mycommands}

\begin{document}

\begin{frontmatter}

\title{The Florence Trigger-Box (FTB) project: an FPGA-based configurable and scalable trigger system}
\author[infnfi]{P. Ottanelli}
\author[infnfi,unifi]{G. Pasquali\corref{cor1}}
\ead{pasquali@fi.infn.it}
\author[infnfi,unifi]{L.Baldesi}
\author[infnfi,unifi]{S.Barlini}
\author[unipd]{R.Bolzonella}
\author[kuleuven]{A.Camaiani}
\author[infnfi]{G.Casini}
\author[infnfi,unifi]{C. Ciampi}
\author[infnlnl]{T.Marchi}
\author[infnfi]{S.Piantelli}
\author[infnfi]{S.Valdrè}
\author[infnct]{G.Verde}
\author[]{    \vskip 0.2cm 
for the NUCL-EX collaboration}

\cortext[cor1]{Corresponding author}
\address[infnfi]{INFN Sezione di Firenze, via G.Sansone 1, 50019 Sesto Fiorentino (FI), Italy}
\address[unifi]{Dipartimento di Fisica e Astronomia, Universit\`a di Firenze, via G.Sansone 1, 50019 Sesto Fiorentino (FI), Italy}
\address[unipd]{Dipartimento di Fisica e Astronomia ``Galileo Galilei'', Universit\`a di Padova, Via F. Marzolo 8, 35131 Padova, Italy}
\address[kuleuven]{Instituut voor Kern- en Stralingsfysica, K.U. Leuven, Celestijnenlaan 200D, B-3001 Leuven, Belgium}
\address[infnlnl]{INFN LNL Legnaro, viale dell'Universit\`a 2, 35020 Legnaro (PD) Italy}
\address[infnct]{INFN Sezione di Catania, via Santa Sofia 64, 95123 Catania, Italy}

\begin{abstract}
A multi-layer trigger system, based on programmable logic devices hosted on VME boards, has been implemented. It is completely scalable and suitable for handling from a few  up to a few hundred channels. Custom software for monitoring and controlling the trigger system has been developed.
\end{abstract}


\end{frontmatter}


\section{Introduction}

 In a trigger-based acquisition system, like those commonly  used in nuclear physics experiments, the trigger system recognizes different kinds of events that can  be handled differently.  It allows both   a reduction of the acquisition data throughput, thus decreasing the dead-time, and  a more efficient use of the available storage. Moreover, it allows for a raw a-posteriori sorting of the acquired data based on the so called trigger pattern, a binary word where each bit is associated with one of the activated triggers. It also makes it possible to selectively acquire events on the base of  their overall multiplicity or by requiring the coincident detection of particles in different parts of the apparatus. For instance, one can acquire and store just a small fraction of the elastic scattering events, which are usually of no interest apart from monitoring and calibration purposes.

The  NUCL-EX collaboration is  in charge of the
GARFIELD+RCo~\cite{Bruno2013} apparatus, installed at the  Laboratori Nazionali di Le\-gna\-ro (LNL) of the Istituto Nazionale di Fisica Nucleare (INFN).
In the last few years, especially after the upgrade of the Front End Electronics (FEE) of GARFIELD+RCo with digitizing electronics~\cite{Pasquali2007126,lnlreportbo}, the need has been recognized for an up-to-date, remotely controlled and configurable trigger system to replace the old one based on legacy NIM modules.
Therefore,  the NUCL-EX collaboration has developed a  general-purpose multi-layer trigger system,  named Florence Trigger-Box (FTB). The FTB  is highly scalable, remotely controlled, and configurable. It has been implemented on programmable logic and is  based on CAEN  V2495 Programmable Logic Units~\cite{CAEN_V2495}.
Using commercially available programming logic units  one can concentrate on writing the ``firmware'', i.e. the logic functions implemented on the programmable logic device. Very limited (if any at all) hardware development is needed. This work is an example of how such boards can be exploited to build a complete trigger system.  Our firmware is implemented in VHSIC Hardware Description Language (VHDL)~\cite{VHDL}, using as a starting point  the template projects provided by CAEN.

 A software code for monitoring and controlling the system has also been developed. It runs under the GNU/Linux operating systems either on a CPU inserted in the same VME bus as the V2495 board, or on a PC connected to VME via a CAEN VME bridge (e.g. CAEN V1718 or V2718 units).

Since its first implementation~\cite{marzocco}, the development of the FTB has been in steady progress  and new features have been added to the project during the last few years. The FTB has also been employed in many experiments, either with GARFIELD+RCo (see, e.g. Refs.~\cite{camaiani2018, Cicerchia_2021}) or with other apparatuses (see, e.g.,  Refs.~\cite{Pasquali2014}~and~\cite{CIAMPI201960}).  
The firmware and the controlling software  have been made available for download~\cite{gitFTB} in the hope that they can be of use to other members of the scientific community. 

In this work, the present version of the firmware will be described. To put it into context, Sec.~\ref{sec:garfield} briefly presents the characteristics of the GARFIELD+RCo apparatus
and the experimental triggers it employs.
Section~\ref{sec:cb_mtb} illustrates the FTB system, i.e. its hardware  (the CAEN V2495 VME board),  its layer structure and the  implementation details of the two layers, and the custom software developed to control the trigger system and monitor its behaviour. Section~\ref{sec:garftrig} shows how the FTB system satisfies the trigger needs of GARFIELD+RCo. Section~\ref{sec:perf} briefly describes the performance of the FTB in terms of dead time and maximum achievable trigger rate.

 \begin{figure}
\center
\includegraphics[width=\textwidth]{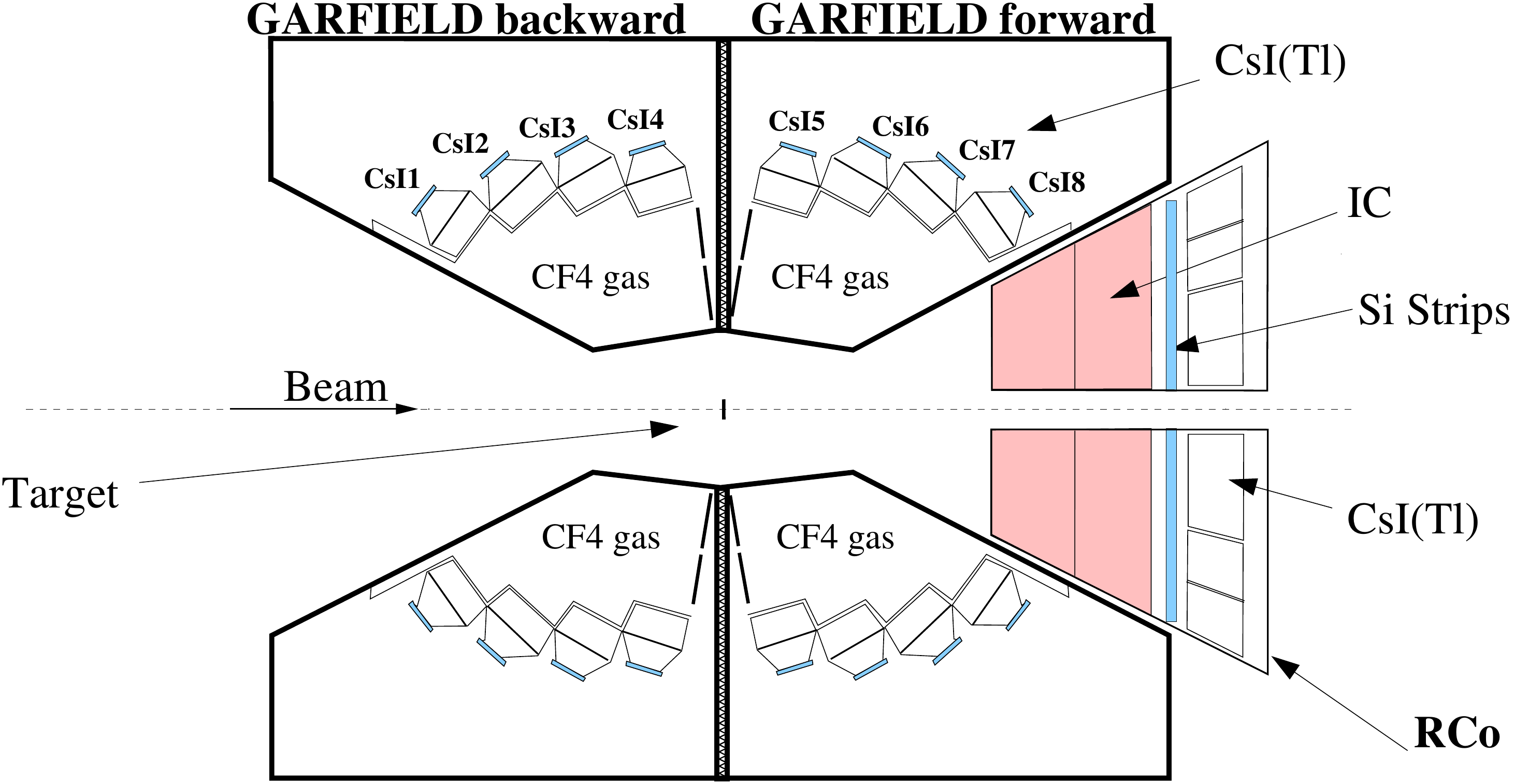}
\caption{Schematic drawing of the GARFIELD+RCo apparatus (adapted from~\cite{Bruno2013}).}\label{fig:garfrco}
\end{figure}

\section{The GARFIELD+RCo apparatus}\label{sec:garfield}

Though the FTB has been thought as a general purpose trigger system, it
has been initially developed for the needs of the GARFIELD+RCo apparatus, a detector array for nuclear fragment detection and identification in heavy ion collisions~\cite{Bruno2013}. In GARFIELD+RCo, nuclear fragments are identified in atomic and  mass number thanks to the so-called $\Delta$E-E technique, i.e. by correlating the energy deposited by the fragment into two detecting layers~\cite{Bromley1962}: the first layer measures an energy loss ($\Delta$E), the second layer stops the particle and measures its residual energy, E.
The layout of the apparatus in shown Fig.~\ref{fig:garfrco}. GARFIELD+RCo  consists of two  parts:

\begin{itemize}
  \item the GARFIELD detector:  a $\Delta$E-E telescope array combining  two Drift Chambers,  for $\Delta$E measurement, and 180 CsI(Tl) scintillators, for residual energy measurement; 
  \item the Ring Counter (RCo):  a segmented three-stage $\Delta$E-E telescope array covering small polar angles.
 \end{itemize}

\noindent 
GARFIELD covers polar angles from $30^o$ to $150^o$ with respect to the beam direction and it  is divided into a ``forward chamber'' (covering polar angles from $30^o$ to $85^o$) and a ``backward'' chamber (covering polar angles from $95^o$ to $150^o$), both segmented into 24 azimuthal sectors (three sectors are actually missing in the backward chamber, to leave room for ancillary detectors). Each sector includes four CsI scintillators, each covering a different polar angle interval. The CsI are placed in such a way as to form rings centered on the beam axis. The rings are numbered from 1 to 4 (backward chamber) and from 5 to 8 (forward chamber). In Fig.~\ref{fig:garfrco} each CsI detector is numbered according to the corresponding ring.
%

\noindent The RCo  covers polar angles from $5^o$ to $17^o$ with respect to the beam direction. It is divided into eight azimuthal sectors, each covering $45^o$ in azimuth. Each sector features three detecting stages:
\begin{itemize}
 \item a Ionization Chamber (IC in Fig.~\ref{fig:garfrco});
 \item a pie-shaped Silicon detector,  segmented into eight annular strips, thus forming eight rings (from which the name ``ring'' counter);
 \item a CsI stage, made up of 6 different crystals,  which  stops and detects the more energetic particles. 
\end{itemize}

\noindent In the RCo, the  $\Delta$E-E technique can be applied either using  the IC vs silicon or the silicon vs CsI(Tl) correlation. Pulse Shape Analysis (PSA) of detector signals is also used for fragment identification and it is applied both to CsI and silicon detectors.
Further details can be found in Ref.~\cite{Bruno2013}.

 \noindent In the present version of the FEE,  all the detectors of the RCo and all the CsI(Tl) of GARFIELD produce  Trigger Request (TReq) signals.
The signals  from the GARFIELD backward chamber and the IC stage of the RCo are digitized with the custom digitizers described in Ref.~\cite{Pasquali2007126}. The TReq from each of these detectors is produced by means of an analog comparator with a programmable threshold. The input signals are treated by a fast filtering stage (a CR-RC filter with 470$\,$ns shaping constant) before being fed to the comparator. It should also be mentioned that the comparator has a low hysteresis and it is therefore affected by some level instability (especially when the trailing edge of the CR-RC signal crosses the threshold level) due to the noise component of the signal.

A newer version of the digitizing FEE handles the signals  from the other detectors, i.e. the GARFIELD forward CsI scintillators and the silicon and CsI detectors of the RCo.
 This version is equipped with a  Cyclone III FPGA and the TReq's are produced by processing the digitized signals with a  bipolar shaper followed by a zero-crossing detector.

All the TReq signals are available on the front connectors of the FEE modules and they are read out by custom electronics, featuring a logic level adapter  to convert from the TTL to the LVDS standard required by the V2495 input.\\

 The following  triggers are commonly used at GARFIELD+RCo during data taking:
 \begin{itemize}
  \item one  trigger for each   GARFIELD chamber (logic sum of the TReq's  from its CsI detectors); they indicate the detection of a particle in GARFIELD;
  \item two  triggers associated with the RCo detector (the logic sum of the Si TReq's and of the CsI TReq's respectively); they indicate the detection of a nuclear fragment at forward angles (useful to select events in which a heavy residue of the nuclear reaction is focused at small polar angles by the kinematics);
\item a  trigger associated with   events produced by a Pulse Generator, for monitoring and calibration purposes;
  \item a  trigger associated with the beam monitors (logic sum of the TReq's  from the beam monitoring detectors); this trigger is usually strongly downscaled (by more than a factor of 50);
  \end{itemize}
  
 Moreover, a coincidence trigger is often used, requiring the presence of one of the GARFIELD triggers and one of the  RCo triggers  within a fixed time coincidence window ( ``GARFIELD \& RCo''  trigger). The coincidence trigger is useful to mark events in which a fragment has been detected by the RCo together with one detected in GARFIELD: such events are good candidates for  inelastic reactions in which at least two particles have been detected. 

 Depending on the experiment, each  trigger is suitably downscaled (or disabled if undesired)   by exploiting a downscale logic and an enabling/disabling trigger mask  (cfr. Sec.~\ref{sec:tboxtop}).

\section{The Florence Trigger Box (FTB)}\label{sec:cb_mtb}

\subsection{The CAEN V2495 board}\label{sec:caen}

The FTB is presently based on the CAEN V2495 board. In the present section, only the features of the VME board which are relevant to this work will be illustrated. We refer to the user manual for further information~\cite{CAEN_V2495}.

The V2495  is a 6U VME board featuring two  FPGAs. One FPGA, the
 ``User FPGA'' (UFPGA, a Cyclone V \cite{altera_cycloneV}), can be programmed by the user.  The \mbox{UFPGA} firmware can be uploaded to an internal EPROM either via the USB interface or via the VME bus. It is then loaded by the \mbox{UFPGA} at power-up. The UFPGA receives the logic signals from the input connectors,  performs the required operations and  drives the 
output signals. 
The second FPGA, the ``Main FPGA'' (MFPGA) 
handles the VME bus interface and the user's access to the \mbox{UFPGA}. Registers and memories 
implemented on the \mbox{UFPGA} can thus be read and written through the VME bus. An internal 50 MHz clock allows for synchronous operation of the internal FPGAs.

Three thirty-two bit connectors (I/O sections A, B and C) form the default I/O implementation of the board, together with two LEMO connectors (I/O section G). 
The input-only A and B connectors accept differential input signals both in LVDS and ECL standards.
The output-only C connector follows the LVDS standard. The G LEMO connectors can serve either as inputs or as outputs -- though not simultaneously -- of either TTL or NIM signals. In  the FTB, one LEMO input is used for an (optional) external VETO signal.

It is possible to expand the I/O capabilities of the unit by adding up to three piggyback expansion boards \footnote{It is thus possible to have up to 128 inputs, which is in fact the configuration used for some of the boards in the GARFIELD+RCo setup.}. Expansion boards featuring 8 NIM-TTL I/O pins, 32 ECL outputs or 32 LVDS inputs are available. Section~\ref{sec:garftrig} illustrates how the  expansions are used in the framework of the GARFIELD+RCo trigger system\footnote{The FTB  code is also available for the (now discontinued) V1495 board~\cite{CAEN_V1495}. The V1495 features a Cyclone UFPGA and a 40$\,$MHz internal clock. The main difference between the V1495 and the V2495 version of the FTB is  the maximum number of inputs: due to the internal capacity of the UFPGA mounted on the V1495, the maximum number of inputs handled by the V1495 version of the FTB is 32.} .

 \begin{figure}
\center
\includegraphics[width=0.7\textwidth]{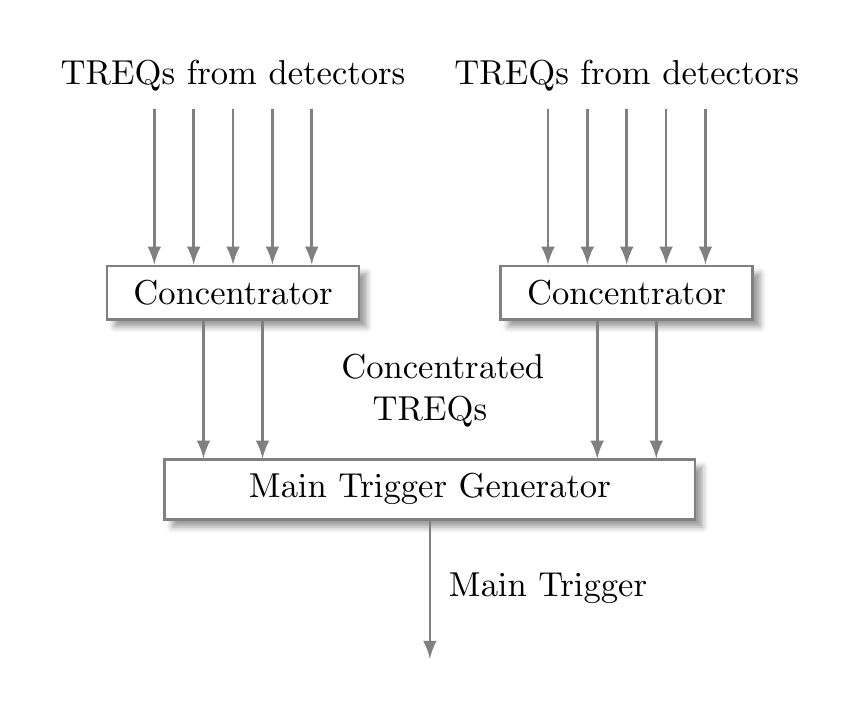}
\caption{Layer structure of the FTB trigger system.}\label{fig:structure}
\end{figure}

\subsection{The layer structure of the FTB}

The FTB trigger system is organized in layers, as shown in Fig.~\ref{fig:structure}, in order to make it easily scalable. A first \textit{concentrator} layer receives the TReq's directly from the single elements (detectors) of the apparatus, grouping them into a smaller number of logic signals (CTReq's, i.e. \textit{Concentrated TReq's}).  For instance, a typical CTReq's could be  the logic sum of the TReq's coming  from detectors of the same type and/or placed at the same polar angle with respect to the beam direction.
The concentrator layer produce also multiplicity triggers, i.e. logic signals related to the number of TReq's received within a given time window.
The second layer,  the heart of the trigger system, produces the Main Trigger (MT) of the experiment by applying the required logic conditions to its logic input signals.  It also handles the dead time by communicating with  the acquisition system.
Accordingly, the FTB  project consists of two different processing layers  (both exploiting  V2495 boards) equipped with two different firmware codes loaded into the \mbox{UFPGA}:

\begin{itemize}
 \item the \textbf{Concentrator Board (CB)} is the first trigger layer, used mainly as a concentrator for trigger requests (TReq's)  from the FEE channels and for building up multiplicity triggers (cfr. Sec.~\ref{sec:cbtop}); 
 \item the \textbf{Main Trigger Board (MTB)} is the last processing layer and it features a more complex structure with respect to the CB (cfr. Sec.~\ref{sec:tboxtop}).
\end{itemize}

\subsection{The Concentrator Board (CB)}\label{sec:cbtop}

The main task of the CB is to combine  the TReq's  from the FEE channels, in order to increase the maximum number of channels that can be managed by the whole system. The CTReq's produced by the CBs are then injected as inputs into the MTB. In addition, the CB produces multiplicity triggers. The TReq signals from the FEE follow the flow illustrated in Fig.~\ref{fig:scheme1}, from left to right. The processing blocks in Fig.~\ref{fig:scheme1} operate as follows:

 \begin{figure}
\hskip -2cm
\includegraphics[width=1.4\textwidth]{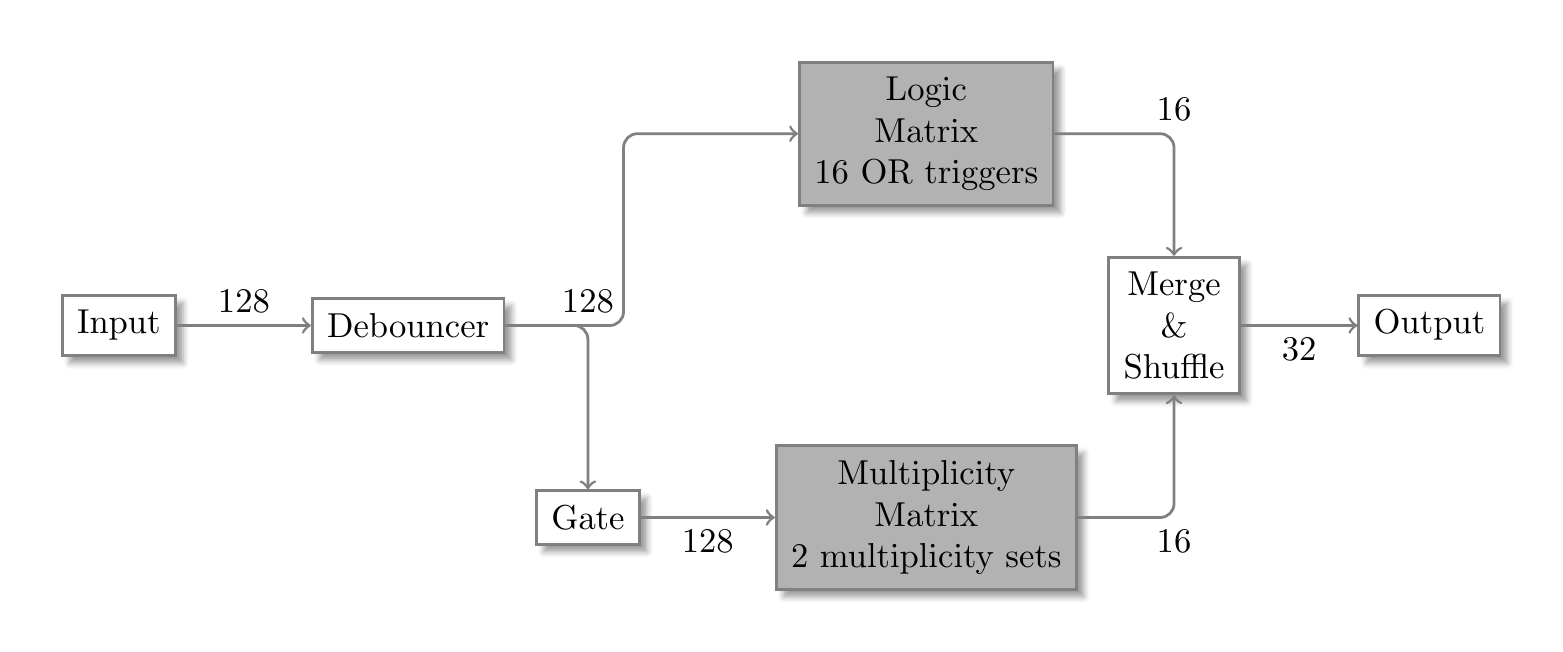}
\caption{Block diagram of the CB firmware. The number of logic signals connecting two adjacent blocks is indicated.}\label{fig:scheme1}
\end{figure}

\begin{itemize}
 \item a 128 input/128 output \textit{Debouncer} cleans the input signals from spurious   transitions and synchronizes them with the internal clock of the V2495 board; the debouncer is needed, e.g., in GARFIELD+RCo, to reject the spurious transitions caused by the short hysteresis of the comparators, as mentioned in Sec.~\ref{sec:garfield};
 \item a 128 input/16 output \textit{Logic Matrix} (\textbf{LM}) performs 16 logic sums (``or'' operation) of the input triggers and generates 16 output signals, each one being the logic sum of a maximum of 128 inputs. For each output, a 128 bit mask allows to select which of the 128 inputs contribute to the logic sum;
 \item on a parallel line, with respect to the LM, a 128 input/128 output \textit{Gate}  transforms the input triggers into fixed length logic pulses; the length of the pulses defines the coincidence window for the multiplicity evaluation and can be set by the user between $1$ and $2^6$ clock periods;
 \item after the gate, a 128 input/16 output \textit{Multiplicity Matrix} (\textbf{MM})   manages two sets of multiplicity triggers. For each set, a mask allows to select the active inputs; each multiplicity set consists of 8 outputs associated with increasing multiplicity values (M$\ge n$, $n=1,\ldots,8$): e.g., the M$\ge 2$ output is set to ``true'' if at least two of the active inputs are ``true'', etc.;

\item the outputs of the two processing lines are  merged by a   \textit{Merge \& Shuffle} block (\textbf{MS}),  to form the full output of the CB, which is 32 bit wide.
 \end{itemize}

 \noindent  To optimize the interconnection between the devices, the MS block divides the 32 output signals into four groups of 8 (groups A-B are LM outputs, groups C-D are MM outputs)
 and the four groups can be re-ordered before being sent to the output connector. The re-ordering capability is particularly useful when using only 16 of the 32 output bits of the CB, as it allows to select which of the outputs has to be associated with the first 16 bits of the 32-bits output connector (slot C of the V2495). 
 
 A summary of the user accessible parameters of the CB, including their possible values,  is presented in Tab.~\ref{tab:CBparams} ($\tau_{clk}$ indicates a clock period). A clarifying example, the application to the GARFIELD+RCo setup, is discussed in Sec.~\ref{sec:garftrig}.

\begin{table}
\caption{User accessible parameters of the CB.}\label{tab:CBparams}
 \begin{center}
\begin{tabular}{|c||c|c|c|}
\hline
Block & Prog.Param. & \#Params & Par.Range\\
\hline
\hline
Logic Matrix &  Masks for OR triggers & $16\cdot N_{inputs}$ & Off(0)/On(1)\\
\hline
Gate &  Width & 1 & $[1,2^6]\tau_{clk}$\\
\hline
Multiplicity Matrix &  Masks for Mult. Sets & $2\cdot N_{inputs}$ & Off(0)/On(1)\\
\hline
Merge \& Shuffle &  Shuffle Output & 1 & \{A,B,C,D\} permutations\\
\hline
\end{tabular}
\end{center}
\end{table}

\subsection{Main Trigger Box} \label{sec:tboxtop}
\begin{figure}
\hskip -2cm
\includegraphics[width=1.4\textwidth]{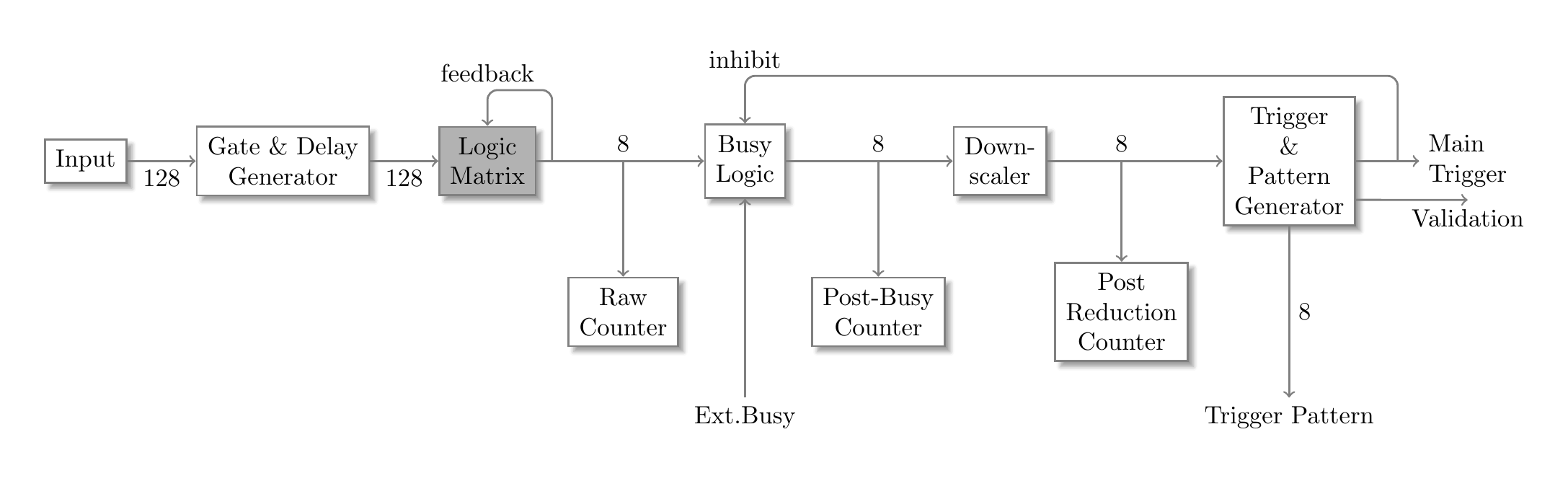}
\caption{Block diagram of the MTB. The Logic Matrix is the only combinatorial block. The signal flow  goes from left to right. The number of logic signals connecting two adjacent blocks is indicated. Two feedback paths are present: the trigger outputs of the Logic Matrix can also be used as inputs for building complex triggers and the Main Trigger is used to inhibit the Busy Logic so that all the triggers firing during the resolving time can be latched in the Trigger Pattern.}\label{fig:fig1}
\end{figure}

The basic structure of the Main Trigger Box, shown in Fig.~\ref{fig:fig1}, is similar to many trigger systems 
employed in nuclear physics experiments. The main inspiration for this trigger box is the trigger system of the FIASCO experiment~\cite{Bini2003497}, which was based on two NIM ``Trigger Box'' units built at GSI, Darmstadt.\\
The main role of the MTB is to combine its input signals (either coming directly from FEE or from the CB) in such a way as to produce the Main Trigger (MT) of the experiment, which is then used to start the acquisition of an event (in GARFIELD+RCo it also starts digital signal processing on the FEE).
The MT signal produced by the MTB results from the logic sum of a few \textit{partial} trigger signals. The partial triggers are obtained from a configurable combination of logic sums and products of up to 128 input signals. Each partial trigger can be downscaled by an integer factor, $n$, so that it contributes to the main trigger only once out of $n$ times.\\


The MTB features 128 logic inputs. Since each CB produces 32 logic outputs, a single MTB can be connected to a maximum of four CBs, so that   up to 512 TReq's can be handled simultaneously\footnote{For larger setups, additional inputs can be managed by using two  CB layers instead of one.}. As shown in Fig.~\ref{fig:fig1}, the processing part of the  MTB consists in five main  blocks:
\begin{itemize}
\item a 128 input/128 output programmable \textit{Gate \& Delay Generator} that latches each input signal 
 when it becomes true and generates a gate signal with user configurable length and delay: though the length is the same for all the inputs, the delay can be set individually, to allow for an optimal time-alignment of the outputs, thus taking into account the different time response of different detectors and/or the different length of the associated transmission lines;
\item a 128 input/8 output programmable \textit{Logic Matrix}: inspired by the Lecroy 8LM CAMAC module, the logic matrix builds up to eight logic sums (logic ``or'' operation) from the 128 inputs\footnote{The number of logic sums could be increased if needed, with just  minimal changes in the VHDL code.}; each input can be logically inverted before the sum and each output can also be inverted, so that the logic product (``and'' operation) is also possible by exploiting the De Morgan law (i.e. $\overline{AB}=\overline{A}+\overline{B}$ with $A$ and $B$ boolean logic values); the outputs of the Logic Matrix constitute the partial triggers whose logic sum, when true, produces the MT; eight feedback paths are provided, thus extending the total number of operands to 136 for each sum; the logic matrix is the only combinatorial part of the MTB (all other blocks being sequential, i.e. clocked)\footnote{Since the Logic Matrix is purely combinatorial, its outputs react in real-time (apart from the unavoidable logic delays) to changes of its inputs. However, in the MTB, input changes are only possible on the rising edge of the clock, since the Gate\&Delay Generator is a sequential block, synchronous with the main clock of the UFPGA.
 };
\item a 8 input/8 output \textit{Busy Logic}: the partial triggers are not passed to the next block if the acquisition system is busy (i.e. during the acquisition dead time); the ``veto'' condition can be set by an external signal connected to one of the LEMO inputs (I/O section G) or it can be automatically set by the MTB after generating a main trigger, in which case the veto can be reset by the acquisition by writing to a register of the MTB;
\item a 8 input/8 output \textit{Downscaler}: its role is to reduce by an integer factor, $n$, the rate at which a partial trigger is passed to the next block: the trigger is passed only once every $n$ times;
\item a 8 input/2 output \textit{Trigger Generator}: it generates the Main Trigger (MT) signal, as the sum of its inputs; the sum is performed after an enabling/disabling 8-bit mask has been applied (the bit mask is programmable); as a matter of fact, the MT  signal is emitted as soon as a partial trigger  becomes true at its input; the state of each partial trigger is then latched in a \textit{Bit Pattern} register after a fixed time interval (\textit{resolving time}, see below).
\end{itemize}

In order to reliably recognize the input trigger request, 
 each input of the the Gate\&Delay Generator is connected to the clock input of a latching flip-flop, i.e.  inputs are not sampled on the fronts of the MTB clock. Therefore, even input signals shorter than the clock period of the UFPGA ($\tau_{clk}=20\,$ns in the V2495) can be reliably recognized. The time length of the Gate\&Delay Generator outputs determines the coincidence time window when the feedback paths of the Logic Matrix are used to produce coincidence triggers, an example being the  ``GARFIELD \& RCo'' partial trigger already mentioned  in Sec.~\ref{sec:garfield}.

 The Bit Pattern filled by the Trigger Generator section can be acquired and stored together with the event data, so that  the partial trigger configuration which actually produced the MT can be retrieved for each event during the data analysis. This allows, e.g., to separate pulser events from the physics during the offline analysis, if a specific partial trigger has been set up for pulser events. Though the MT is produced as soon as a partial trigger becomes true at its input,
  partial triggers arriving late with respect to the first one  can still be recognized and latched in the Bit Pattern register, provided that they arrive within a fixed Resolving Time, i.e. a programmable time window starting when the Main Trigger becomes true.  In fact, the Busy Logic block allows its input signals to pass until the end of the resolving time, so that  partial triggers can still go through. The status of the Bit Pattern is latched at the end of the resolving time.
 The MT signal stays active during  the whole resolving time (programmable from 1 to $2^6$ clock cycles).

 A \textit{validation} (VAL) signal is also produced by the Main Trigger section at the end of the Resolving Time (the length of the VAL signal is programmable from 1 to $2^{16}$ clock cycles).   The VAL signal could be subjected to further conditions, so that one could use the  MT as a conversion/digitization start signal and the VAL as an acquire/store enabling signal.
The absence of a true VAL could then act as a ``fast clear'' condition in cases where the event should be rejected.  For instance, in the GARFIELD+RCo FEE, each channel starts processing its input signal on receiving the MT. After a few microseconds, it tests for the presence of a true VAL signal and resets itself if VAL is false\footnote{At the moment, the  ``fast clear'' possibility is not exploited in GARFIELD+RCo: in fact, for each MT signal, a VAL signal is always produced.}.
 
 Apart from the MT and the VAL signal, the output signals of the MTB also include the  partial triggers (after downscaling) and a serial output emitting the value of the bit-pattern for each event\footnote{The latter is used in GARFIELD+RCo in order to ``inject'' the bit-pattern value into the current event data. Since a I/O register module is not available for the FAIR bus on which the present acquisition system~\cite{Ordine1998} is based, the most cost effective solution is to emit the Bit Pattern value from the MTB in serial form, then digitizing and decoding its value using one of the FEE channels. In a VME based acquisition, on the other hand, the Bit Pattern register can be accessed directly from the VME bus.}.
The output signals of the MTB can be redirected either to slot C (LVDS standard) or to slot F (ECL standard or NIM/TTL). If a NIM/TTL piggyback board is plugged into slot F, only the relevant signals (Main Trigger, Validation and Bit Pattern) are sent to the 8 output connectors and the outputs follow the NIM logic standard.

\begin{table}
\caption{User accessible parameters of the MTB.}\label{tab:MTBparams}
 \begin{center}
\begin{tabular}{|c||c|c|c|}
\hline
Block & Prog.Param. & \#Params & Par.Range\\
\hline
\hline
Gate\&Delay Gen. & Width  & 1&$[1,2^6]\tau_{clk}$ \\
\hline
Gate\&Delay Gen. & Delay & $N_{input}$ & $[1,2^6]\tau_{clk}$ \\
\hline
 Logic Matrix & 8 Trigger Masks & $8\cdot N_{input}$ & Off/On(+/-)\\
 \hline
  Logic Matrix & 8 Feedback Masks & $8\cdot 8$ & Off/On(+/-)\\
 \hline
  Logic Matrix & Output Status & $8$ & Off/On(+/-)\\
 \hline
 Downscaler &  scaling factors  & 8 & 1 to $2^{16}-1$\\
 \hline
Trigger Gen. &  Active Triggers  & 8 & Off/On\\
 \hline
 Trigger Gen. &  Width/Delay  & 2 &  $[1,2^6]\tau_{clk}$ \\
 \hline
 Trigger Gen. &  Resolving Time  & 1 &  $[1,2^6]\tau_{clk}$ \\
 \hline
\end{tabular}
\end{center}
\end{table}
 
The bottom row of Fig.~\ref{fig:fig1} shows three 32-bit counters which are used to monitor the dead time of the acquisition and to acquire the actual count rate (useful, e.g., for determining cross sections). Each counter is actually eightfold, i.e. a counter is provided for each partial trigger. The leftmost counter  (\textit{Raw Counter}) is incremented each time a partial trigger is produced by the Logic Matrix (one counter for each trigger). The \textit{Post-Busy Counter} is incremented only when the trigger passes the veto-logic, i.e. triggers happening during dead-time are not counted. The \textit{Post-Reduction Counter} is incremented by triggers passing through the Downscaler, i.e. actually contributing to the Main Trigger. In GARFIELD+RCo, the acquisition system periodically reads and resets the 24 counters, calculating and displaying the rate for each partial trigger pre- and post-busy and post-reduction. The percentage of dead time is
given by

$$
f_{dead}\,=\,\frac{N_{Raw}-N_{Post-Busy}}{N_{Raw}}\,\times\,100\,\%
$$

\noindent where $N_{Raw}$ and $N_{Post-Busy}$ are the counts registered by the Raw Counter and by the Post-Busy Counter respectively.

 A summary of the user accessible parameters of the MTB, including their possible values,  is presented in Tab.~\ref{tab:MTBparams}.

\subsection{Common Features and Monitoring}\label{sec:common}
In order for the whole system to be fully scalable and flexible, both the CB and the MTB are equipped with some extra features concerning the input stage.
Specifically, the additional piggyback boards mounted on slots D and E are automatically recognized and the effective number of inputs is evaluated depending on the configuration. For the 8 channels NIM/TTL piggyback boards the logic standard of the input can be selected using a specific register which is accessible from the control software.
\newline
For the debugging and monitoring of the whole system, each board is equipped with a \textit{Logic Analyzer} that consists of:
\begin{itemize}
 \item an input multiplexer that allows to choose between different preset groups of 32 internal logic signals to be stored;
 \item a 32-bit wide circular buffer (capable of storing 2048 32-bit words) which is constantly updated on each clock edge with the output values  of the multiplexer;
 \item a 4096 word, 32-bit wide memory that is filled with values coming from the circular buffer whenever a Logic Analyzer Trigger (LAT) condition is verified (see below); this memory can be accessed via the VME bus (cfr. Sec.~\ref{sec:la});
 \item a LAT generation  logic to start the acquisition of  the multiplexed values. 
\end{itemize}
The condition for the filling of  the Logic Analyzer memory can be either a software trigger (i.e. a trigger generated by the controlling software, cfr. Sec.~\ref{sec:sw}) or a  true value of the Main Trigger (for the MTB) or a true value of a logic sum calculated from  a selection of the 32 input values (a selection mask is provided and can be written by the controlling software).\\
Like in a digital oscilloscope, the acquired values are divided into a pre-trigger (pre-LAT) and a post-trigger (post-LAT) portion: up to 2048 pre-LAT values (i.e. logic input values immediately preceding the LAT condition) can be acquired. The actual  pre-LAT length can be set by the controlling software. The maximum number of acquired values is 4096 (81.92$\,\mu$s at a 50$\,$MHz clock).

\subsection{Controlling software}\label{sec:sw}

The controlling software is written in C++ programming language, also exploiting the ROOT~\cite{ROOT} graphical libraries (the GUI window inherits from the ROOT  TGMainFrame class). It can affect the behaviour of both the CB and the MTB by writing to their memory mapped control registers.  
Information such as the board model (V2495 or V1495), the firmware type (CB or MTB) and the effective number of inputs, the latter depending on the installed piggyback boards, can  be easily retrieved by the software at start-up. The Graphical User Interface  of the software automatically adapts to the connected board. For instance, when the software handles a CB, the input gate and delay section is not present and there is no way to invert an input. The controlling software can run either on a VME CPU or on a PC (under the GNU/Linux operating system), the latter connected to the VME bus through a CAEN VME bridge.
More details can be found on the git site~\cite{gitFTB}.
%
%


\subsection{Monitor software (Logic Analyzer) and multiplexer}\label{sec:la}


The monitor software exploits the Logic Analyzer section of the CB and of the MTB. It allows the user:

\begin{itemize}

\item to select which signals have to be stored and displayed;

\item to change the  pre-LAT length (from 1 to 2048 $\tau_{clk}$) and the overall  length (from 1 to 4096 $\tau_{clk}$);

\item  to select the signals used to trigger the LA acquisition (any one of the 32 displayed signals plus the MT);

\item to  enable/disable the data acquisition and display.

\end{itemize}

The state (high/low) of each signal is plotted as a function of time and updated after each trigger (just like in an ordinary oscilloscope). 
Using the different MUX presets it is possible to inspect all the intermediate signals of the trigger box, thus allowing for a fine adjustment of the parameters, particularly the input delays of the MTB which are critical parameters for obtaining correct coincidences between triggers generated in the same event.

\section{The GARFIELD+RCo trigger system}\label{sec:garftrig}

In GARFIELD+RCo, the TReq's  from the GARFIELD+RCo detectors are handled by a single concentrator layer.
The  number of TReq's in GARFIELD+RCo is $\sim300$. Therefore,  the concentrator layer of the  trigger system employs three CBs, each one connected to a different section of the apparatus. A single MTB, receiving the outputs of the three CBs and also other sources (e.g. the trigger output of the pulse generator), produces the MT which starts the event acquisition. Overall, four CAEN V2495 boards (three CBs and a MTB) are used.
Going into some more detail:
\begin{itemize}
 \item the \textbf{Garf\_BW} CB is the concentrator associated with the backward chamber, it is configured to accept up to 96 inputs (only 84 presently used for the backward chamber);  it produces four outputs  that are each the logic sum of the CsI TReq's for a given CsI ring; it also produces a global  CsI TReq (the logic sum of all the rings); its output includes 8 multiplicity triggers, i.e.  M$\ge n$, with $n=1,\ldots,8$, where $n$ is the number of active CsI TReq's  (e.g. the M$\ge2$ multiplicity signal becomes true when two or more CsI TReq's are true);
 \item the \textbf{Garf\_FW} CB is the concentrator associated with the forward chamber, and it's configured the same way as the former (except all the 96 inputs are used);
 \item the \textbf{RCo} CB handles all the 120 TReq's  from the RCo detector; it produces  the logic sum of the IC TReq's, the logic sum of the Silicon TReq's (actually 8 signals, a logic sum of the TReq's for each ring) and the logic sum of the CsI TReq's, together with two   multiplicity sets, one for the silicon stage and one for the CsI stage. 
 \end{itemize}
 The output of the \textbf{RCo} CB is fed into the input slot A of the MTB, and the two 16-bit outputs of the \textbf{Garf\_BW} and \textbf{Garf\_FW} CBs are connected to the input slot B. To feed extra TReq's to the MTB, an additional eight input NIM/TTL piggyback board is added on slot D. Other  TReq's, such as those coming from two fast plastic scintillators used for beam monitoring and cross section normalization, are inserted into the system by exploiting the free inputs of \textbf{Garf\_BW}. 


 \section{Dead time and performance}\label{sec:perf}
 
The dead time (DT) is the time for which an incoming TReq is ignored by the trigger system, since it is busy serving a previous request. Different blocks of the CB and MTB firmware feature different DT characteristics. 

\subsection{CB performance}

The CB performance is limited by the pulse width of the debouncer output, which is presently   160$\,$ns. In terms of the well-known DT models~\cite{Knoll2010}, the CB debouncer is paralyzable (i.e. a new TReq arriving at an already active input will add further 160$\,$ns to the output pulse duration). The counting rate of a single input corresponding to 1$\,$\% DT is thus about 63$\,$kHz. This is greatly in excess of the maximum counting rate achievable at GARFIELD+RCo, which is presently constrained to much lower values (less than 1$\,$kHz)  by the signal digitization and processing time of the FEE.  
In principle, the output pulse width of the debouncer could be reduced to 40$\,$ns thus allowing for a 250$\,$kHz counting rate at 1$\,$\% DT. When a debouncer is not actually needed, the same performance could be obtained by removing the debouncer  (i.e. by connecting the inputs directly to the LM and the \textit{Gate} blocks, see Fig.~\ref{fig:scheme1}), thus also changing the CB behaviour from a paralyzable to a nonparalyzable one.

The multiplicity trigger section of the CB, based on a coincidence logic, features a nonparalyzable behaviour. The coincidence time is determined by the width of a ``gate'' generator and it is adjusted according to the requirements of each experiment. At the typical setting of 200$\,$ns, the total counting rate (i.e. the overall rate considering all inputs) corresponding to 1$\,$\% DT is 50$\,$kHz.

\subsection{MTB performance}

The counting rate performance of the MTB depends on its internal settings, especially if coincidence triggers are built within its Logic Matrix. The DT of a single channel of the Gate\&Delay Generator is given by the width of its output signal plus its delay time. The DT model is nonparalyzable: for zero delay time and a 200$\,$ns time width, the counting rate at 1$\,$\% DT is 50$\,$kHz. It could be increased up to 250$\,$kHz by reducing the width to its minimum (40$\,$ns), if compatible with the experimental conditions. However, the Logic Matrix puts a more stringent limitation, since it works according to a paralyzable model. In a pure ``logic sum'' configuration (i.e. no coincidence required) its count rate at 1$\,$\% DT is 50$\,$kHz (250$\,$kHz) for a 200$\,$ns (40$\,$ns) input pulse width.

All the reported counting rates, for both the CB and the MTB, would scale by about a factor of ten if evaluated at a 10$\,$\% DT.

 \section*{Conclusions}
 
The Florence Trigger Box (FTB), a compact trigger system based on commercially available VME boards, has been developed within the Italian NUCL-EX collaboration and it has been described in this work. The system is general purpose, easily configurable, and scalable. Control and monitor software allows for handling the trigger system remotely during the experiment.
The FTB includes a set of scalers for dead-time evaluation and cross section normalization.
Complex trigger conditions can be implemented with no  need of additional cabling, thanks to the internal logic matrix and its ``feedback'' paths.
Being based on programmable logic (FPGAs), the system is highly integrated and easily upgradable. For instance, the number of partial triggers could be increased or different types of experimental triggers could be generated according to different triggering conditions.

The FTB 
has also been employed in smaller setups, where just one V2495 board is capable of handling all the trigger requirements (i.e. in the acquisition of prototype SiC detectors~\cite{CIAMPI201960}, to build coincidences between the different stages of the $\Delta$E-E telescopes). As an example of the versatility of the system, the new features which need to be implemented in view of the new acquisition system of GARFIELD+RCo currently under development, will not require new hardware, just some VHDL programming.

The VHDL code and all the needed files are available for download~\cite{gitFTB}.

\section*{Acknowledgements}

The authors would like to thank G.Marzocco for starting the FTB development as part of his  bachelor degree thesis. We are also grateful to M.Bini for his skilful assistance during the MTB development and to A.Olmi and G.Poggi for the fruitful discussions during the preparation of this work. The assistance of CAEN SpA, and in particular of Luca Colombini, is also gratefully acknowledged.

\bibliography{./gab_bibtex} 
\bibliographystyle{elsarticle-template/elsarticle-num}


\end{document}

%% file: mycommands.tex
 \definecolor{Gainsboro}{rgb}{0.86, 0.86, 0.86}

\newcommand{\begmyitemize}{\begin{itemize}[<+-| alert@+>]}
\newcommand{\begmyitemizeB}{\begin{itemize}[<+->]}

\newlength{\parskipbackup}
\newlength{\parindentbackup}
\makeatletter

\def\donotresetequations{{%
    \let\@@elt\relax
    \def\@elt##1{%
        \expandafter\ifx\csname ##1\endcsname\c@equation%
        \else%
            \@@elt {##1}%
        \fi%
    }%
    \edef\beamer@overlaycounterresets{\beamer@overlaycounterresets}%
    \let\@elt\relax%
    \def\@@elt{\@elt}%
    \xdef\beamer@overlaycounterresets{\beamer@overlaycounterresets}%
}}

\makeatother
